\nonstopmode

\documentclass[prl,twocolumn]{revtex4}
\usepackage{graphicx}
\usepackage{natbib}
\usepackage{epstopdf}
\usepackage[hidelinks, colorlinks = true, urlcolor=blue, linkcolor=blue, citecolor= blue]{hyperref}
\usepackage{color}
\usepackage{amsmath}
\usepackage{lettrine}
\begin{document}

\preprint{}

\title{ Hydro-Responsive Curling of the Resurrection Plant\\ {\textit{Selaginella lepidophylla}}}
\author{Ahmad Rafsanjani$^1$, V\'eronique Brul\'e$^2$, Tamara L. Western$^2$, Damiano Pasini$^1$}
\email[Correspondence to ]{damiano.pasini@mcgill.ca}

\affiliation{$^1$Mechanical Engineering Department, McGill University, 817 Sherbrooke Street West, Montr\'eal, QC, H3A OC3, Canada\\ $^2$Biology Department, McGill University, 1205 Avenue Docteur Penfield, Montr\'eal, QC, H3A 1B1, Canada}

\date{\today}

\hyphenpenalty=5000
\begin{abstract}

The spirally arranged stems of the spikemoss {\it Selaginella lepidophylla}, an ancient resurrection plant, compactly curl into a nest-ball shape upon dehydration. 
Due to its spiral phyllotaxy, older outer stems on the plant interlace and envelope the younger inner stems forming the plant centre.
Stem curling is a morphological mechanism that limits photoinhibitory and thermal damages the plant might experience in arid environments. Here, we investigate the distinct conformational changes of outer and inner stems of {\it S. lepidophylla} triggered by dehydration. Outer stems bend into circular rings in a relatively short period of desiccation, whereas inner stems curl slowly into spirals due to hydro-actuated strain gradient along their length. This arrangement eases both the tight packing of the plant during desiccation and its fast opening upon rehydration. The insights gained from this work shed light on the hydro-responsive movements in plants and might contribute to the development of deployable structures with remarkable shape transformations in response to environmental stimuli.

\end{abstract}

\maketitle

Resurrection plants are vascular plants tolerant to extreme vegetative desiccation that are able to resume normal growth and metabolic activity upon rehydration~\citep{Rascio05}.
The spikemoss {\it Selaginella lepidophylla} is an ancient~\citep{Banks09} resurrection plant native to Chihuahuan desert (Mexico and United States) that shows dramatic curling and uncurling with changes in plant hydration~\citep{Eickmeier80}.
When dehydrated, the spirally arranged stems of {\it S. lepidophylla} tightly curl to form a rough sphere.
As a result of this morphology, the outer stems serve to substantially reduce solar radiation ($>99.7\%$) exposed to inner stems at the centre of the plant~\citep{Eickmeier86}.
The morphological and anatomical traits of {\it S. lepidophylla} in relation to the curling of its stems were examined at the turn of the twentieth century~\citep{DuSablon1888}.
At that time, it was elucidated that the movements of the tissues are entirely physical -rather than biophysical- and depend upon the hygroscopic capacities of the tissues~\citep{Uphof20}.

While fascinating for botanists, adaptive movements in plants can inspire material scientists and engineers to exploit the underlying mechanisms for the development of innovative biomimetic materials and actuating devices that show intriguing shape transformations in response to environmental stimuli~\citep{Ionov13, Erb13, deHaan14, Zhao14, Guiducci14}.
Nastic movements in plants are generally driven by hydration motors of osmotic, colloid or fibrous design, where the direction of the movement is determined by integrated features of mobile tissue, rather than by stimulus direction~\citep{Stahlberg09, Burgert09}.
In particular, the plant motion in fibrous motors relies on the relatively slow variation of the water content within the internal capillary spaces of parallel-arranged cellulose fibres in the cell walls~\citep{Skotheim05}. 
Changes in hydration results in anisotropic swelling and shrinkage strains which emerge in the direction transverse to the long axis of the cells.
By combining different layups of cellulose layers, plants generate a wide variety of water-controlled actuators which can trigger a diverse range of complex movements.
The swelling/shrinkage induced movements are strongly dependent on the stresses generated through moisture uptake which may also occur in non-living remnants of plants.
Thus, sometimes they can be viewed from a purely mechanical perspective~\citep{Bertinetti13}.
Opening of pine cones~\citep{Dawson97}, water responsive movements of the skeleton of the desert plants {\it Anastatica hierochuntica}~\citep{Friedman78, Hegazy06} and {\it Asteriscus pygmaeus}~\citep{Gutterman94}, unfolding mechanism of the seed capsules of the desert ice plants from {\it Aizoaceae} family~\citep{Parolin06, Harrington11}, self-burial mechanism of hygroscopically responsive {\it Erodium} awns~\citep{Evangelista11, Abraham12, Aharoni12}, twisting of seed pods~\citep{Armon11} and the walk and jump of {\it Equisetum} spores~\citep{Marmottant13} are examples of plant movements in response to environmental stimuli.

In this work, we investigate the moisture responsive curling of the stems of the spikemoss {\it S. lepidophylla} through a multidisciplinary combination of experiments, theory and numerical simulations.
We aim at understanding the underlying mechanisms of nastic movements in this resurrection plant as a paradigm for the design of novel mechanisms of water-controlled actuation and structural deployment.

\section*{\large Results}

\begin{figure*} [!t]
\centering
\includegraphics [width=0.9\textwidth]{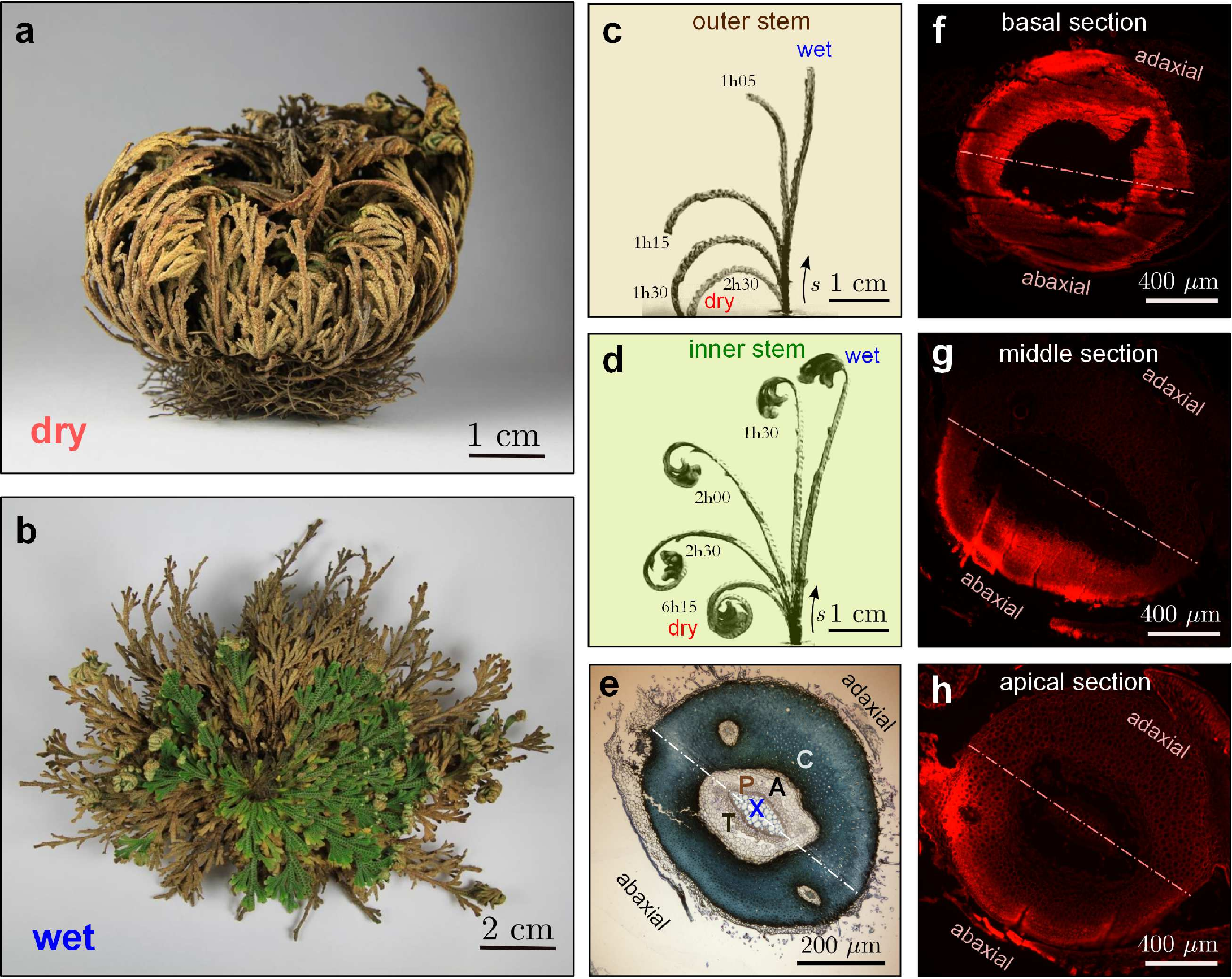}
 \caption{  Morphology and composition of the resurrection plant \textit{Selaginella lepidophylla}. A plant in ({\bf a}) desiccated and ({\bf b}) hydrated states. 
{\it S. lepidophylla} exhibits a spiral phyllotaxy with the youngest stems near the centre of the plant and the oldest stems near the outermost edge of the plant.
Curling sequence of fully hydrated ({\bf c}) outer and ({\bf d}) inner stems of {\it S. lepidophylla} during dehydration.
Both stems exhibit large deformation with distinct curling patterns.
({\bf e}) Cross section of a toluidine blue O (TBO) stained, Spurr's resin-embedded inner stem  (basal region) showing the location of the cortex (C), xylem (X), phloem (P), and specialized trabeculae cells (T) in the air space (A) separating the protostele from the cortex.
While lignified cell walls (dark blue) are found throughout the inner zone of the cortical tissue, there appears to be a difference of cell density between the adaxial and abaxial zones.
Basic Fuchsin was used to visualize lignin within the cell walls of the stem tissue.
The cross-sections of a Basic Fuchsin stained, paraffin-embedded, inner stem at ({\bf f}) the basal section reveal lignification (bright red) in all the ground tissue, whereas ({\bf g}) the middle of the stem, shows lignified tissues in the abaxial side of the stem.
This staining pattern is reduced to a narrow strip at ({\bf h}) the apical tip of the stem.}
\label{Fig1}
\end{figure*}

{\bf Plant morphology}. 
In the hydrated state, the spirally arranged stems of {\it S. lepidophylla} are flat and outstretched and, upon dehydration, they compactly curl into a spherical nest-ball shape with an average diameter of 6 to 8 cm (Fig.~\ref{Fig1}a and b and Supplementary Movie 1). 
The exterior stems are typically fragile and have a grey-brown colour. 
Protected within these older outer stems may be several layers of inner stems forming the centre of the flattened, green rosette of the plant that uncurl and are photosynthetically active upon rehydration in their native state~\citep{Eickmeier86}.

{\bf Shape transformation of stems}. 
The curling and uncurling of isolated stems taken from outer and inner portions of the spiral rosette of {\it S. lepidophylla} were filmed during hydration and dehydration.
Fig.~\ref{Fig1}c and d respectively show the curling sequences of outer and inner stems in the course of dehydration from the initial wet condition to the final dry state. 
When hydrated, both inner and outer stems were flattened, but when dehydrated, they exhibited large deformation with distinct curling patterns.
Upon dehydration, the outer stems bend into an approximately circular ring in a relatively short period of desiccation, whereas the inner stems curl slowly and form a spiral configuration~(Supplementary Movie 2).

\begin{figure*} [!t]
\centering
\includegraphics [width=0.95\textwidth]{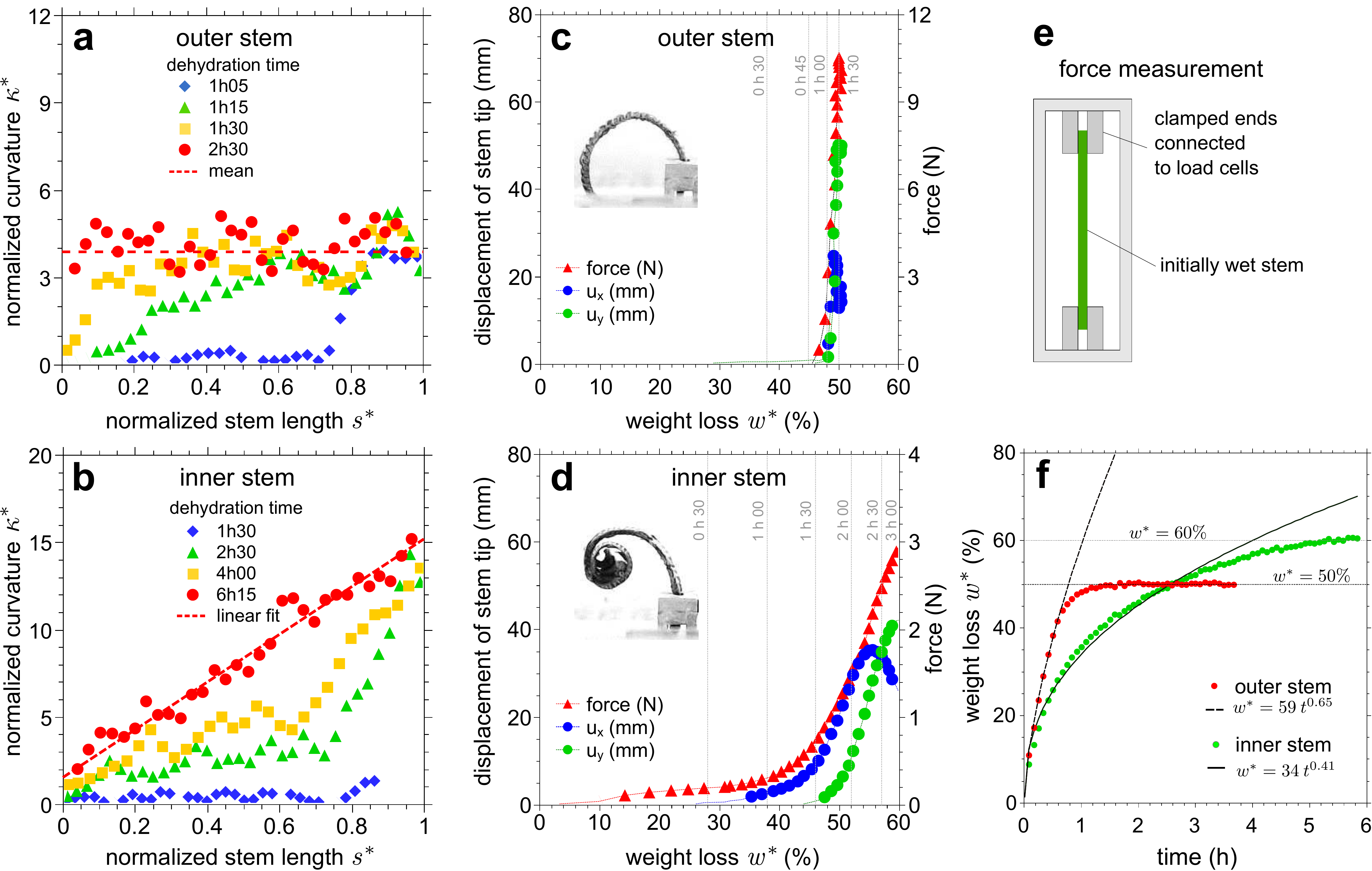}
 \caption{ Response of outer and inner stems to dehydration.
The normalized curvature $\kappa^*=\kappa l$ as a function of the normalized arc-length $s^*=s/l$ of (a) outer and (b) inner stems of {\it S. lepidophylla} at different time intervals.
Mechanical response of initially wet stems to dehydration for (c) the outer and (d) the inner stems.
Plotted as a function of the weight loss $w^*$ of the stem during dehydration, the absolute displacements of the stem tip, $u_x$ and $u_y$, and the reaction force of a similar stem clamped at its both ends illustrate that stem curling correlates with induced-stress buildup.
The vertical lines in gray represent the associated dehydration time.
(e) force measurement setup.
(f) Dehydration weight loss $w^*$ of outer and inner stems as a function of time.
The early drying behaviour before reaching an equilibrium is fitted with a power-law function.}
  \label{Fig2}
\end{figure*}

{\bf Morphology and composition of stem sections}.
The observation of curling mechanisms of outer and inner stems naturally leads to the question of whether their distinct curling patterns are correlated to the morphology and composition of their underlying tissue. 
The stem of {\it S. lepidophylla} is composed of an annular-shaped layer of cortical tissue and an inner, central vascular bundle (protostele), which is separated from the cortical layer by an air-filled canal (Fig.~\ref{Fig1}e).
The vascular tissue is comprised of a central bundle of xylem cells that is surrounded by a sheath of phloem cells (i.e., an amphicribral bundle)~\citep{Brighigna02}.
Specialized cells, termed trabeculae, are found within the air-filled canal. 
Due to their thick, cellulosic cell walls, trabeculae provide structural support, bridging the airspace between the central vascular bundle and the cortical tissue~\citep{Harholt12}.

The composition of the cell walls was investigated by staining Spurr's resin-embedded stem sections (Supplementary Information S1) with Toluidine Blue O (TBO) for the detection of lignin and pectin under bright field microscopy.
Fig.~\ref{Fig1}e shows a cross-section of an inner stem at the root-base interface, and it reveals that, while lignified cell walls (dark blue) are  found throughout the inner cortex tissue, there is an asymmetry in cell density, whereby cells appear to be smaller and/or denser on the abaxial side of the cross section.
Further investigation with basic Fuchsin staining of paraffin-embedded sections (Supplementary Information S2) allowed detection of variation of lignified tissues at different sections along the stem length.
Consistent with the TBO results, lignification is found throughout the inner cortex in the basal sections~(Fig.~\ref{Fig1}f), whereas the middle of the stem shows lignified tissues in the abaxial side of the stem~(Fig.~\ref{Fig1}g).
The lignified tissues are reduced to a narrow strip at the apical tip of the stem~(Fig.~\ref{Fig1}h).
For outer stems, the staining pattern is similar to that seen at the basal section of inner stems without significant variation along the stem length (Fig.~\ref{Fig1}e and f; data not shown).
The spatial variation of lignification along the stem length reduces from the base to the apical tip of inner stems of {\it S. lepidophylla}, and it may play a crucial role in their spiralling curling mechanism.

{\bf Curvature characterization of stems.}
The variation of the curvature along the length of outer and inner stems (see Fig.~\ref{Fig1}c and d) was characterized at different time intervals from an analysis of the discrete curvatures~\citep{Langer05} of the stem centreline~(Fig.~\ref{Fig2}a and b and Supplementary Information S3). 
The stems had a small natural twist which we neglect in this analysis.
The curvature and stem length are normalized with the total length $l$ respectively as $\kappa^*=\kappa l$ and $s^*=s/l$.
Starting from an almost straight shape, after one hour of drying, the outer stem started to curl with a progressive curling front which was initially close to the tip before propagating toward the base.
The final dehydrated outer stem resembles an arc with a constant curvature.
A propagating curling front was also observed for the inner stem in the course of dehydration where its apical tip exhibited more mobility and compliance compared to its base.  
The maximum achievable curvature of the inner stem increases linearly along the stem length, i.e. $\kappa\propto s$ representing the geometry of an Euler (Cornu) spiral.

{\bf Mechanical response of stems to dehydration}. 
We investigated the mechanical aspects of the water-controlled curling of outer and inner stems in more detail.
The absolute displacements of the apical tip ($u_x$ and $u_y$) of the outer and inner stems where tracked during drying~(Fig.~\ref{Fig2}c and d).
Both the stems experience large deformation.
Furthermore, fully hydrated stems were clamped at both ends and the reaction forces that prohibited stem curling were measured~(Fig.~\ref{Fig2}e).
In a given period of time during dehydration, the weight loss $w^*$ of the wet stems was also determined (Fig.~\ref{Fig2}f), a measure that allowed to translate dehydration time into weight loss.

The mechanical response of each stem was strongly dependent on its moisture content.
For the outer stem, a switch-like behaviour was observed.
At early times of dehydration, the mobility of the stem tip was not notable and the generated axial force in the restrained stem was very small.
Once the outer stem reached a certain weight loss level, i.e. $w^*\simeq45\%$, its movement accelerated and its tip moved downward.
At the same water content, a very sharp increase in the force response of the restrained stem was observed.
On the other hand, the inner stem responded smoothly to dehydration.
Initially, the stem remained still until $w^*\simeq35\%$, then a moderately uniform transition was observed in both displacement and restraining force profiles.
The change of the weight associated to the main vertical displacement of the apical tip for the outer stem was $\Delta w^*\simeq5\%$, whereas $\Delta w^*\simeq25\%$ for the inner stem.
The maximum vertical velocity of the tip ($du_y/dt$) of the outer stem ($\simeq2.7$ mm/min) during dehydration was threefold larger than the one of the inner stem~($\simeq0.9$ mm/min).
Also, the resultant restraining force during dehydration in the outer stem was about threefold larger than the force in the inner stem.
The outer stem reached its equilibrium weight loss, i.e. $w^*=50\%$, after 1 hour whereas for the inner stem it took about 5 hours to reach equilibrium, i.e. $w^*=60\%$.
The drying rate of the outer stem was faster than the inner stem.
A power law behaviour for the early drying rate before reaching the equilibrium was observed with an exponent of 0.65 for the outer stem and 0.41 for the inner one, suggesting a diffusive transport mechanism.

\begin{figure*} [!t]
\centering
\linespread{1.0}
\includegraphics [width=0.9\textwidth]{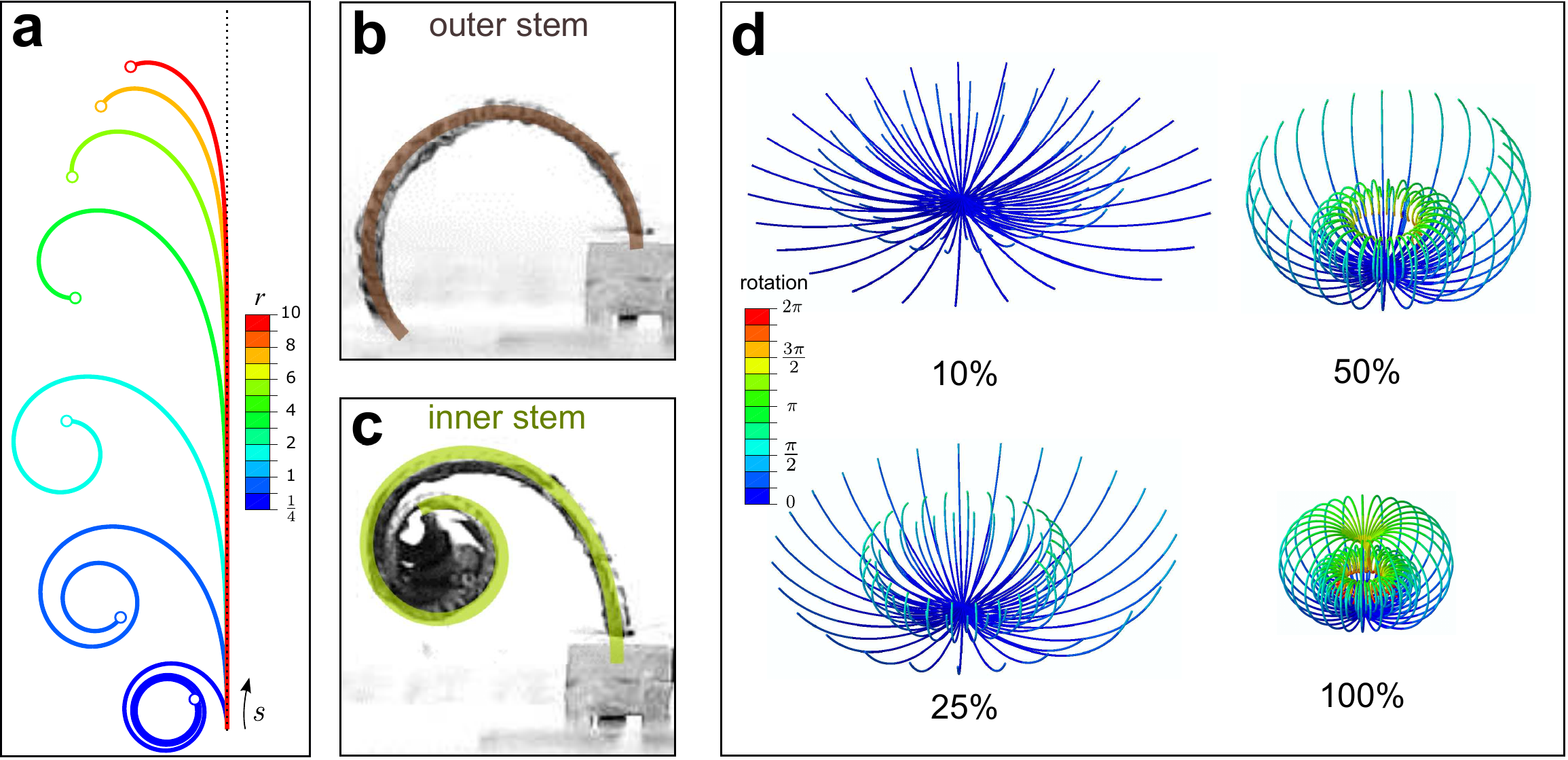}
 \caption{ Curling mechanisms of inner and outer stems.
 (a) The curling patterns of virtual stems with power-law hydro-actuation strain gradient ($\kappa=a^{r+1} s^r$) for different $r$ exponents when $\eta=24$.
 The dotted line shows the initial undeformed state.
 The deformed shape of the dehydrated (b) outer and (c) inner stems of {\it S. lepidophylla} are reproduced using large deformation FE simulations by setting $\eta=\kappa^*$ and are overlaid (solid lines) on stem images.
The curvature of the non-inner stem model is constant ($\kappa^*=3.93$), whereas for the inner stem model, the curvature varies linearly with the stem length ($\kappa^*=1.6+13.64\: s^*$).
(d) A simplified FE model for hydro-responsive curling of {\it S. lepidophylla} plant showing the cooperative packing of outer and inner stems at different stages of dehydration. 
The contours represent the absolute rotation of the stems in radian.
Note that in actual {\it S. lepidophylla} plants, the spiral phyllotaxy would yield a gradient of curvature moving from the centre of the plant.
Also, both the properties of the inner stems, and their shading by the more rapidly closing outer stems will slow their dehydration process, thus the timing of the movement of the inner stems will be delayed in real plants.}
  \label{Fig3}
\end{figure*}

{\bf Curling mechanisms of inner and outer stems}. 
To elucidate the spiralling pattern of inner stems, we adopted a geometrical model based on the normalized Euler (Cornu) spiral, which is defined as a curve with curvature $\kappa$ that linearly changes with its curve length $s$ i.e. $\kappa=a^2 s$.
In this setting, the Cartesian coordinates of a normalized Euler spiral shape are given by the following Fresnel-like integrals:
\begin{equation}
\label{eq:1} 
x(\tilde s)=\sqrt\frac{\pi}{a} \int_0^{\tilde s} \sin \frac{\tau^2}{2} d\tau,\quad
y(\tilde s)=\sqrt\frac{\pi}{a}  \int_0^{\tilde s} \cos \frac{\tau^2}{2} d\tau
\end{equation}
where $\tilde s \in [0,a]$ is the curve parameter and $a \in [0,\eta]$ represents a desiccation factor, where $a=0$ indicates the initial fully hydrated state and $\eta$ the hydro-actuation strain gradient generated in the stem section upon dehydration. 
Generalization of the above geometrical model to a hypothetical class of spirals with power-law curvature (defined by the Ces\`aro equation $\kappa=a^{r+1} s^r$) allows us to investigate the role of material variation along the stem length (Fig.~\ref{Fig3}a and Supplementary Information S4). 
The mobility of the stems decrease by increasing the exponent $r$, an outcome that indicates the curling pattern reflects the hydro-actuation capacity of the tissue.
Alternatively, stem spiralling can be simulated via finite element (FE) modelling of inextensible elastic beams by introducing hydro-actuation strain gradient through the beam cross section~(Fig.~\ref{Fig3}b and c).
The curling of spirally arranged stems of {\it S. lepidophylla} triggered by dehydration has been examined with a simplified FE model of the whole plant~(Fig.~\ref{Fig3}d).
The curling of exterior stems with constant curvature allows them to envelope spiralled inner stems in the centre of the plant. 

\begin{figure*} [!t]
\centering
\includegraphics [width=0.9\textwidth]{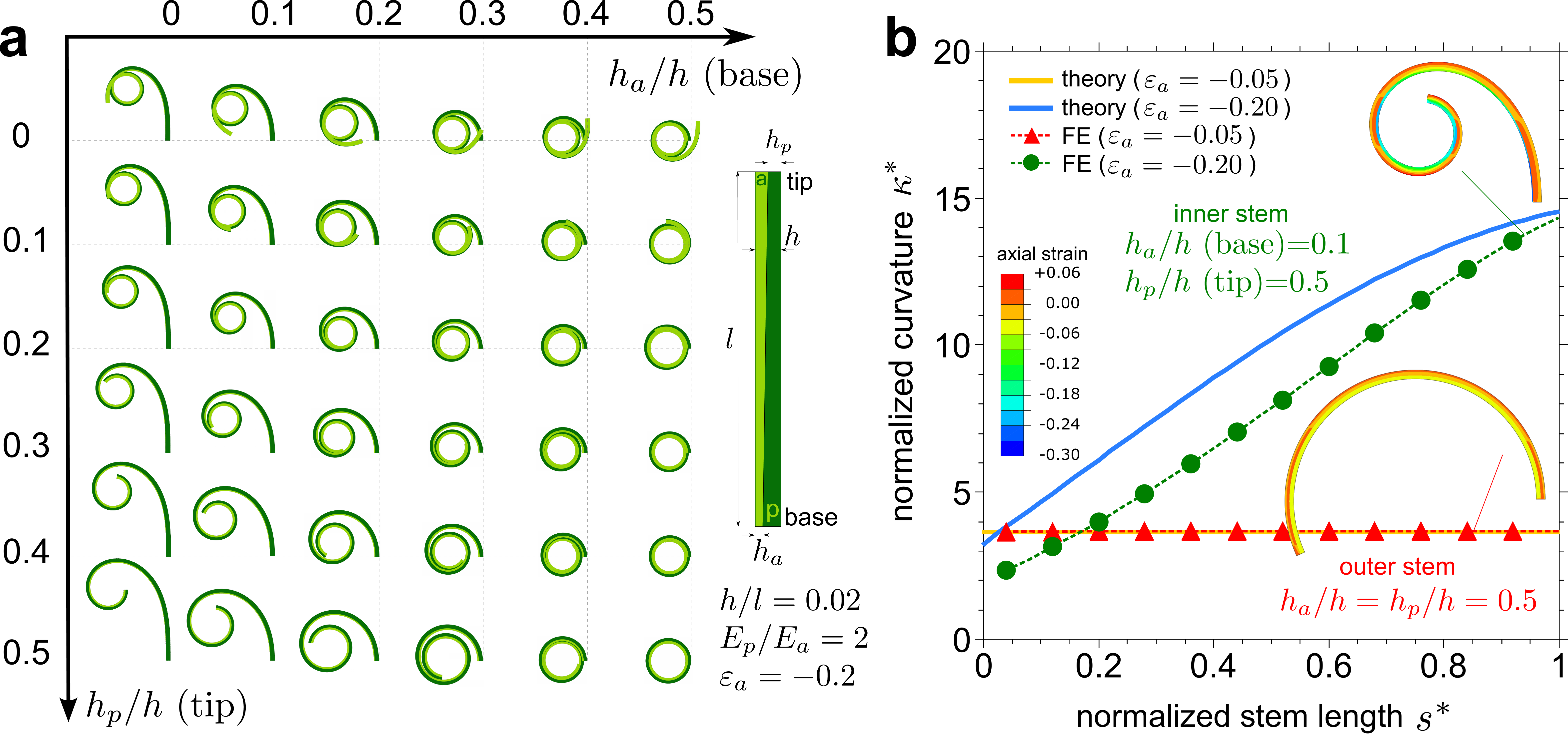}
 \caption{ Curling pattern of bilayer stems.
(a) Conformations of bilayer stems ($h/l=0.02$) composed of elastic active ({\it a}) and passive ({\it p}) layers ($E_p/E_a=2$) subjected to an actuation contraction strain $\varepsilon_a=-0.2$ obtained with large deformation FE simulations for layer profiles with linearly varying thickness along the stem length.
(b) Normalized curvature $\kappa^*$ of bilayer models  of an outer stem with a constant relative thickness ($h_a/h=h_p/h=0.5$) and an inner stem with a linearly varying relative thickness ($h_a/h=0.1$ at the base and $h_p/h=0.5$ at the tip of the stem) under action of shrinkage strains ($\varepsilon_a=-0.05$ for the outer stem and $\varepsilon_a=-0.2$ for the inner stem) in the active layer.
Results obtained with theoretical Timoshenko model (Eq.~(\ref{eq:2})) and FE simulations.
The inset shows the corresponding axial strain contours of deformed stems.}
\label{Fig4}
\end{figure*}

From structural point of view, the stem of {\it S. lepidophylla} can be seen as a bilayer composed of non-lignified active (a) and lignified passive (p) layers that curls upon dehydration due to a mismatch in the eigenstrains developed in each layer.
In outer stems, the relative thickness of the constituent layers is constant, whereas in inner stems the relative thickness of the active (passive) layer increases (decreases) from the base to the apical tip of the stem~(Fig.~\ref{Fig4}a and Fig.~\ref{Fig1}f-h). 
The curvature of a bilayer in terms of elastic modulus, geometry and actuation eigenstrains ($\varepsilon_a$ and $\varepsilon_p$) of the constituent layers is given by the Timoshenko~\citep{Timoshenko25} bimetallic model as:
\begin{equation}
\label{eq:2}
\kappa =\frac{6(1+m)^2 (\varepsilon_a-\varepsilon_p)/h}{3(1+m)^2+(1+mn)(m^2+1/mn)}
\end{equation}
where $h=h_a+h_p$ is the total thickness of the bilayer with thickness ratio $m=h_p/h_a$ and elastic moduli ratio $n=E_p/E_a$. 
The conformations of several bilayer stems ($h/l=0.02$, $E_p/E_a=2$) subjected to an actuation contraction strain ($\varepsilon_a=-0.2$) where the constituents thickness varies linearly along the stem length are computed by large deformation FE simulations (Fig.~\ref{Fig4}a).
In particular, when $h_a/h$ ratio increases at the base of the stem, a transition from spiral to circular configuration is observed.
In addition, for increasingly higher values of $h_p/h$ ratio, the number of turns of each stem concepts reduces during deformation.
The curvature of representative bilayer models of an outer stem with constant relative thickness and an inner stem with linear thickness profile is calculated by Eq.~(\ref{eq:2}) and FE simulations~(Fig.~\ref{Fig4}b).
The FE model account for geometric non-linearities that large actuation strains trigger, as Fig S3 in the supplemental information further explains, while the Timoshenko bimetal model is only accurate for small deformation.
The results compare well with the deformed shape of the stems, and the predicted curvature follows similar trends observed in experiments.
From this analysis, we can gather that the proportion of the constituent passive to active tissues along the stem length governs the resulting shape of the stem.
 
\section*{\large Discussion}

Stem curling triggered by desiccation is a morphological mechanism in the desert plant {\it S. lepidophylla}.
This movement has an ecophysiological importance that limits photoinhibitory and thermal damage to the plant and provides a way to overcome bright-light, high-temperature, and water-deficit stresses~\citep{Lebkuecher93}.
We observed distinct large deformation mechanisms (Fig.~\ref{Fig1}c and d), mechanical responses (Fig.~\ref{Fig2}a-d) and time scales (Fig.~\ref{Fig2}f) in moisture responsive curling of outer and inner stems of {\it S. lepidophylla} which may ameliorate their photo-protection function.
Our study reveals that two mechanisms are potentially effective~(see Fig.~\ref{Fig3}d) in moisture responsive curling of the spikemoss {\it S. lepidophylla}:
(i) stiff outer stems located in the exterior portion of the plant behave as classical bilayers curling with an approximately constant curvature, and (ii) compliant inner stems located in the centre of the plant introduce a class of rod-like hygromorphs~\citep{Reyssat09} with planar spiralling capacity.
This arrangement, when paired with the plant's spiral phyllotaxy that yields a gradient of stem curling patterns, can adaptively regulate the packing and deployment of the whole plant by according the closure of the plant in dry periods and its opening at times of water retrieval.

One of the main differences between inner and outer stems of {\it S. lepidophylla} lies in their shape transformation mechanisms. 
In both cases, the hydro-responsive movements are mechanical and governed by the properties of the tissues and chemical composition of the cell walls. 
At the level studied, the main difference affecting bending in the outer stems appears to be the presence of asymmetric cell density between the abaxial and adaxial sides of the stem. 
For inner stems, however, microscopic studies also revealed the presence of an asymmetric lignification in the cortical tissue towards the abaxial side of the stem~(Fig.~\ref{Fig1}e).
Further, in inner stems, the distribution of lignified cells varies along the stem length~(Fig.~\ref{Fig1}f-h).
The presence of the lignified cells alters the hydro-actuation capacity of the stem and locally increases the stiffness of the cortical tissue.
Upon dehydration, differential shrinkage strains are induced in the tissue, and are relaxed through conformational changes of the stems (Fig.~\ref{Fig1}c and d).
The spatial hydro-actuation strain gradient along the stem length plays an important role in the spiralling pattern of dehydrated stems (Fig.~\ref{Fig3}a).
Conformations similar to those of the inner stems have been observed in the growth process of coiling tendrils~\citep{Wang13}.
In the balloon vine tendril {\it Cardiospermum hallachum}, gelatinous fibres with highly lignified secondary walls generate a contractile force that converts elongated, straight tendrils into a planar spiral shape~\citep{Bowling09}.

The stems of {\it S. lepidophylla} can robustly curl/uncurl over several dehydration/hydration cycles~(Supplementary Movie 3) without structural damage.
The reversibility of the movement indicates that the shape of the stems mirrors their water content.
The stems respond to dehydration once the water loss exceeds a certain threshold~(Fig.~\ref{Fig2}c and d).
This behaviour can be attributed to the existence of free water in liquid form in the interior of the cells and bound water within the fibrous cell walls, a phenomenon also observed in wood cells~\citep{Siau84}.
During dehydration, free water molecules can easily leave the cell cavities (though perhaps more slowly from the cytoplasm if cells are living), and despite their significant volume, they appear to only have minor influence on the deformation of the stem (Fig.~\ref{Fig2}c and d). 
On the other hand, the bound water molecules are strongly absorbed into the microscopic voids of the fibrous cell walls.
During drying desiccation, the fibrous cell walls shrink and produce the driving force for deformation, thereby leading to stem curling.
Differences in the degree of curling and the rate of water loss can be affected by the lignification status of cell walls, as cellulose can hydrogen-bond water molecules and lignin is hydrophobic.

In summary, the spikemoss {\it S. lepidophylla} exploits simple but effective strategies to carefully pack and deploy many of its rod-like stems~(Fig.~\ref{Fig3}d and Supplementary Movie 1).
The spiral phyllotaxy is organized outward in order of ascending length and increasing lignification along the length of the stems(Fig.~\ref{Fig1}b).
This arrangement, in cooperation with the identified curling mechanisms~(Fig.~\ref{Fig3}b and c), facilitates tight packing and fast opening of the plant with a minimized interlocking between stems despite their large deformations.
The insight gained from this study might set the stage for the design and development of deployable structures (e.g. in communication antennas and morphogenetic architectural design concepts) and actuating devices that can yield programmable shape transformations~(e.g. stent deployment devices) in constant feedback and interaction with their functioning environment.

\section*{\large Methods}
\linespread{1.0}
\small

{\bf Plant materials}. 
Mature dehydrated {\it S. lepidophylla} were obtained from Canadian Air Plants in New Brunswick, Canada. Plants were maintained unplanted in a desiccated state under laboratory room conditions ($25^\circ$C, 50\% relative humidity). 

{\bf Time-lapse video capture}. 
Time-lapse videos were captured using the Logitech C920 HD Pro Webcam (1080p, Carl Zeiss optics) and Video Velocity Time-Lapse Studio software (Candylabs). 
{\it S. lepidophylla} plants were allowed to rehydrate for 24 hours, and individual inner stems (chosen from 1/4 to 1/3 of the way along the spiral to match the length of mature stems) and outer stems (chosen from approximately 3/4 the way along the spiral) of $\sim$5 cm length were cut from the plant at the root-stem interface. 
Microphylls (leaves) were cut off and the stem was secured in a metal clamp which was affixed to the base of a square Petri dish. 
Stems were then allowed to air-dry for approximately 6 hours, during which time changes in their curvature were captured via time-lapse filming at frame rate of 1 min$^{-1}$. 
Stems were taken from three different {\it S. lepidophylla} plants. 
In total, four inner stems and four outer stems were tested and displayed similar curling/uncurling patterns. 
The models presented in this article were based upon a single representative stem from both inner and outer sample pools. 

{\bf Drying force measurement}.
Rehydrated stems were stripped of microphylls and secured at both ends by clamps and the drying induced tensile force was measured using an ADMET MicroEP machine. 
Stems (8 inner and 8 outer) were then allowed to air-dry and the change in load over time was calculated every five minutes for a total of 220 minutes. 

{\bf Weight measurement}.
The change in weight between rehydrated and dehydrated states of stems was calculated using a Mettler AJ100 analytical balance. 
Stems were rehydrated overnight and excess water was blotted from the surface of the stem. Individual outer and inner stems were placed upon the balance and allowed to air-dry for 220 and 350 minutes, respectively. 
The weight loss percentage was calculated with respect to the weight of the stem at its fully hydrated state as $w^*=100\times(w_{wet}-w)/w_{wet}$.

{\bf Staining and microscopy}.
Toluidine Blue O (TBO) used for the detection of lignin and pectin. 
Spurr's resin-embedded sections were stained with 0.05\% TBO in 0.01M PO$_4$ buffer (pH 5.7) for 10s on a hotplate ($60^\circ$C) and were rinsed with deionised water. 
Prepared sections were observed with bright field microscopy.

Basic Fuchsin was used to confirm and detect lignin~\citep{Dharmawardhana92}. 
Paraffin-embedded sections were stained in 0.0001\% Basic Fuchsin (in 70\% ethanol) for 5 minutes, washed in 70\% ethanol for 2 minutes, and rinsed briefly in deionised water. 
Samples were mounted in 50\% glycerol and observed with the TX2 (RFP filter set).
Prepared sections were viewed using a Leica DM6000B epifluorescence microscope.

{\bf Finite element simulations}.
Finite element (FE) simulations of simplified outer and inner stem models performed using the commercial package ABAQUS (Rising Sun Mills, Providence, RI, USA).
The hydro-actuation strain is modelled with thermal expansion, where temperature represents moisture content.
The rod-like stems are modelled as inextensible elastic beams and are discretised with hybrid quadratic beam elements (B32H).
The curling is induced by introducing thermal strain gradient through the beam cross section.
The spatial variation of hydro-actuation capacity along the stem length is introduced into the model using an analytic field which is derived from curvature profiles.

The FE bilayer model is discretised with elastic plane stress quadratic elements (CPS6).
The stem is clamped at its base.
The geometrical nonlinearities are taken into account, but the self-contact of the stem is neglected. 
The curling of the stems is simulated by imposing a negative thermal strain in the active layer.
A mesh size sensitivity analysis is performed and based on that a mesh with about 7 elements along the stem thickness ($\sim$5000 elements) was chosen. 
This choice leaded to fairly consistent results in the range of the parameters considered in this work and enabled us to smoothly extract curvature values during the post-processing stage. 
Further information on the FE model of bilayer stems is provided in Supplementary Information S5.

\section*{Acknowledgement}
Ahmad Rafsanjani is recipient of Early PostDoc Mobility Fellowship from \textit{Swiss National Science Foundation} (SNSF) and acknowledges the financial support provided under grant no. 152255. 
Financial support from \textit{le Fonds de Recherche du Qu\'ebec-Nature et Technologies} (FRQNT) under grant no. 173224 is greatly acknowledged.

\section*{Author Contributions}

A.R., V.B. T.L.W. and D.P. conceived the problem and designed research.
V.B. and A.R. conducted experiments. 
A.R. performed simulations.
All authors are participated in interpretation of results, manuscript writing and commenting.

\section*{Additional Information}

{\bf Supplementary Information} accompanies this paper.

{\bf Competing financial interests}: The authors declare no competing financial interests.


\newpage
\setcounter{figure}{0}
\setcounter{equation}{0}

\makeatletter
\renewcommand{\thefigure}{S\@arabic\c@figure}
\renewcommand{\thetable}{S\@arabic\c@table}
\renewcommand{\theequation}{S\@arabic\c@equation}
\makeatother

\section{SUPPLEMENTAL MATERIALS}

\subsection{Preparation of Spurr's~resin-embedded sections}

Inner and outer stem sections were fixed in 3\% glutaraldehyde in 0.1M PO$_4$ buffer (pH 7.0) for 16 hours at $4^\circ$C on a nutator. 
Samples were then washed $3\times10$ minutes in 0.05M PO$_4$ buffer. 
Samples were post-fixed in a 1\% OsO$_4$ solution (in 0.05M PO$_4$ buffer) for 2 hours at room temperature. 
Samples were rinsed in deionised water and subjected to an ethanol series (10 minutes in 30\% ethanol, 1 hour each in 50\%, 70\% and 85\%, 95\% and 2x 100\% ethanol) at room temperature. 
Samples were washed in 100\% propylene oxide for 30 minutes, after which they were changed into a mixture of propylene oxide and Spurr's resin (2 hours each of 3:1 PO:Spurr's, 1:1 PO:Spurr's, and 1:3 PO:Spurr's). 
Samples were left in 100\% Spurr's overnight at room temperature. 
The next day, samples were changed into fresh Spurr's twice and left overnight. 
Samples were changed into fresh Spurr's the next day and were polymerized in open tubes at $60^\circ$C for 48 hours.
Samples were sectioned using a Leica EM UC6 Ultramicrotome and were mounted on regular glass slides.

\subsection{Preparation of paraffin-embedded sections}

Inner and outer stem sections were fixed in FAA (4\% paraformaldehyde, 5\% acetic acid and 50\% ethanol) for seven days at $4^\circ$C on a nutator. 
The FAA solution was changed for fresh FAA on day 3 of fixation. 
Stems were then washed twice for 30 minutes each in 50\% ethanol $4^\circ$C and left overnight in 70\% ethanol at $4^\circ$C on a nutator. 
Samples were dehydrated through an ethanol series (1hr each 85\%, 95\%, $2\times$ 100\%) at room temperature. 
After dehydration, samples were transferred to a mixture of xylene and ethanol (1hr each: 75\% ethanol:25\% xylene, 50\% ethanol:50\% xylene, 25\% ethanol:75\% xylene, $2\times$ 100\% xylene). 
Samples were placed in fresh xylene in scintillation vials and paraplast chips were added ($1/4$ volume of xylene) overnight at room temperature. 
The next day, the vials were placed at $42^\circ$C for 30 minutes. 
Another $1/4$ volume of paraplast chips was added and samples were incubated at $60^\circ$C for 6 hours. 
Paraplast:xylene solution was discarded and replaced by molten paraplast (i.e. paraplast chips that had been melted at $60^\circ$C for 24 hours). 
Samples were then left at $60^\circ$C for one week and the molten paraplast was changed for fresh solution twice a day. 
Samples were embedded in fresh paraplast in small Petri dishes and the paraplast was allowed to harden at room temperature overnight. 
Samples were sectioned using a Leica RM2245 Microtome and were mounted on positively charged slides. 
Samples were deparaffinised prior to staining with 2x 100\% xylene (15 minutes each).

\subsection{Discrete curvature characterization}
\label{S3}

\begin{figure} [ht]
\centering
\includegraphics [width=\columnwidth]{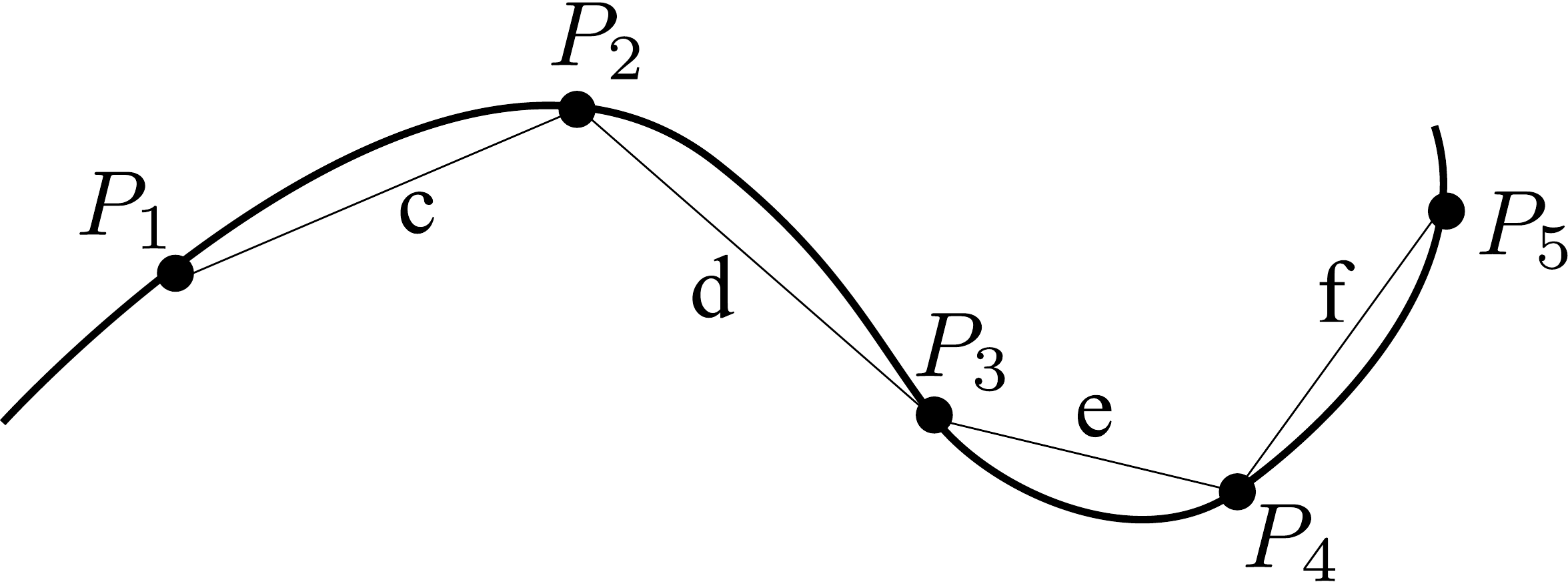}
 \caption{ Discrete representation of a smooth curve.}
\label{Fig_S1}
\end{figure}

The curvature of the stems of {\it S. lepidophylla} (Fig. 2a and b) are characterized by accurate estimations of the curvature of a smooth curve from its discrete approximation.
Fig.~\ref{Fig_S1} shows a segment of a smooth curve which is represented by a polyline with five points $P_1$ to $P_5$, with the corresponding edges $\overrightarrow{P_iP_{i+1}}$ ($i=1,2,3,4$) denoted by $\boldsymbol{c}$, $\boldsymbol{d}$, $\boldsymbol{e}$ and $\boldsymbol{f}$, and their lengths are $c$, $d$, $e$ and $f$.
The curve can be represented by a Taylor series expansion~\citep{Langer05}.
A linear approximation for the true curvature vector $\boldsymbol{\kappa}$ can be obtained by finite difference approach and if all edges have equal length, the convergence is even quadratic.

\begin{equation}
\label{eq:S1}
\boldsymbol{\kappa} =\frac{2}{d+e}\left (\frac{\boldsymbol{e}}{e}-\frac{\boldsymbol{d}}{d} \right)
\end{equation}

\subsection{Geometrical model based on Euler spiral}

Towards rationalizing the spiralling behaviour of the living stems, we adopt a geometrical model based on the definition of the normalized Euler (Cornu) spiral.
By definition, an Euler spiral is a curve whose curvature $\kappa$ changes linearly with its curve length $s$, i.e. $\kappa=a^2 s$ where $a$ is a constant.
This definition can be generalized to consider the role of material variation along the stem length. 
For a general class of power-law curvature (e.g. induced by the functionally graded hydro-actuation capacity of the tissue) defined by the Ces\`aro equation $\kappa=-a^{r+1} s^r$, the parametric equations for the spiral profile read:

\begin{widetext}
\begin{equation}
\label{eq:S2}
x(\tilde s)=\frac{\tilde s^{2+r}}{(1+r)(2+r)\sqrt[r+1]{a}} {}_1F_2\left(\left\{\frac{1}{2}+\frac{1}{2(1+r)}\right \};\left\{\frac{3}{2},\frac{3}{2}+\frac{1}{2(1+r)} \right \};-\frac{\tilde s^{2(1+r)}}{4(1+r)^2} \right)
\end{equation}

\begin{equation}
\label{eq:S3}
y(\tilde s)=\frac{\tilde s}{\sqrt[r+1]{a}} {}_1F_2\left(\left\{\frac{1}{2(1+r)}\right \};\left\{\frac{1}{2},1+\frac{1}{2(1+r)} \right \};-\frac{\tilde s^{2(1+r)}}{4(1+r)^2} \right)
\end{equation}
\end{widetext}
where $\tilde s\in [0,a]$, $a\in[0,\eta]$ and ${}_pF_q(a_p;b_q;z)$ is the generalized hypergeometric function~\citep{Askey10}.
These equations were evaluated in the computational software Mathematica (Wolfram) using the built-in function \texttt{HypergeometricPFQ}.

The above simplified model is used to investigate the role of material variation along the stem length as illustrated in Fig.~3a in the article.

\subsection{Finite element model for bilayer stems}
Finite element (FE) simulations of bilayer stem models performed using the commercial package ABAQUS 6.11 (SIMULIA, Rising Sun Mills, Providence, RI, USA). A Python script is written to systematically create stem models.

\begin{figure} [ht]
\centering
\includegraphics [width=0.8\columnwidth]{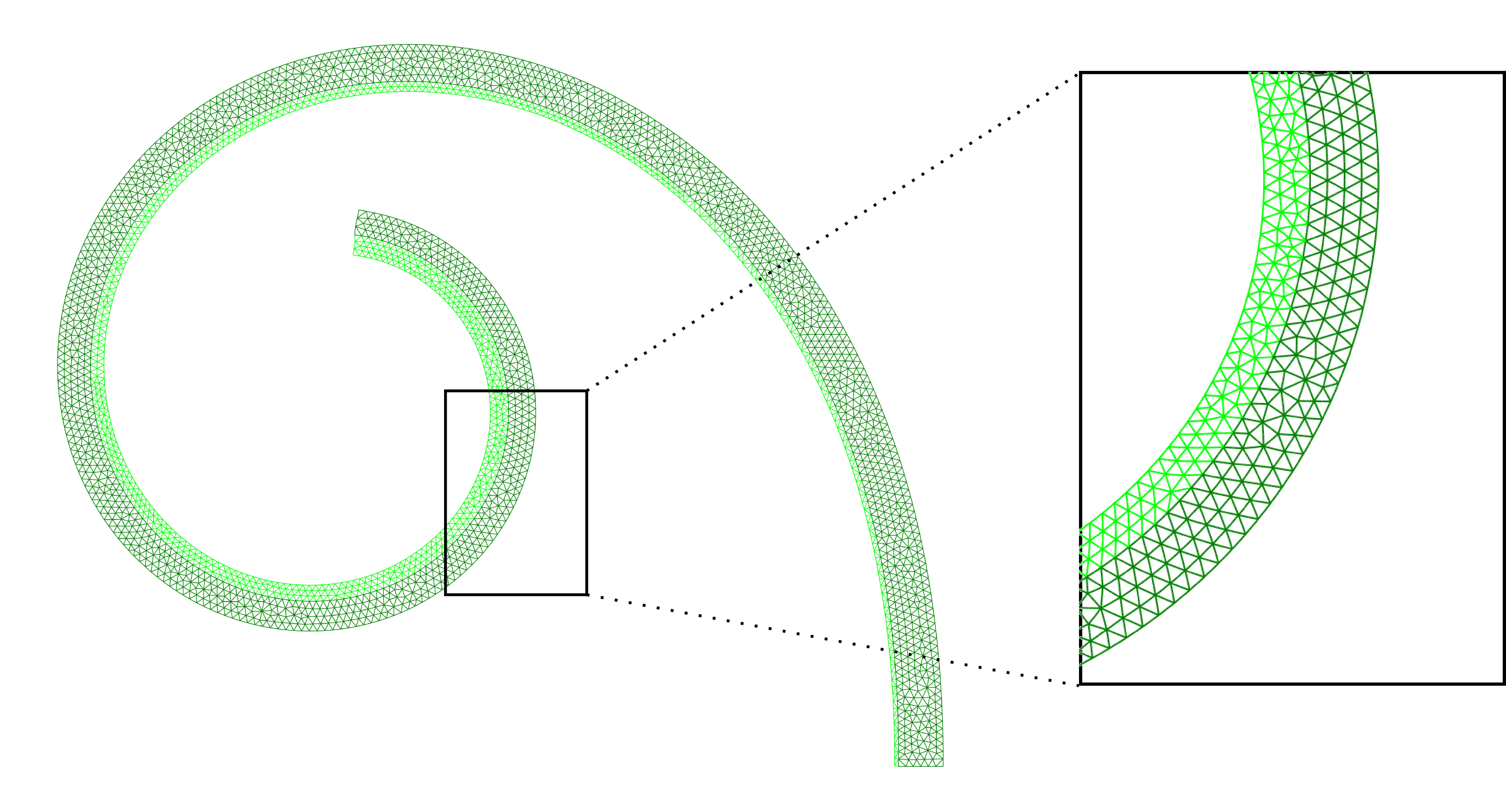}
 \caption{  Finite element mesh for a bilayer stem.
A bilayer stem ($h/l=0.02$, $h_a/h=0.1$ at base and $h_p/h=0.5$ at tip) is meshed with 4857 triangular plane stress quadratic elements (CPS6) in ABAQUS.}
\label{Fig_S2}
\end{figure}

The bilayer is composed of a soft active (a) and a stiff passive (p) elastic layer, which have respectively the elastic moduli of $E_a$ and $E_p$ and the actuations strains of $\varepsilon_a$ and $\varepsilon_p$.
The Poisson's ratio for both layers is $\nu_a=\nu_p=0.3$.
The hydro-actuated strain is modelled with thermal expansion, where temperature represents moisture content.
The stem is clamped at its base.
Geometric nonlinearities are taken into account by activating \texttt{NLGEOM} option in ABAQUS which allows for large-deformation analysis.

A mesh size sensitivity analysis is performed and based on that a mesh with about 7 elements along the stem thickness ($\sim$ 5000 triangular plane stress quadratic elements, CPS6) gives consistent results in the range of the parameters considered in this work.
Fig.~\ref{Fig_S2} shows the mesh for a bilayer stem in its deformed state.
This mesh resolution allows us to characterize curvature smoothly along the stems centreline following the procedure introduced in~\ref{S3}.

\begin{figure} [b]
\centering
\includegraphics [width=\columnwidth]{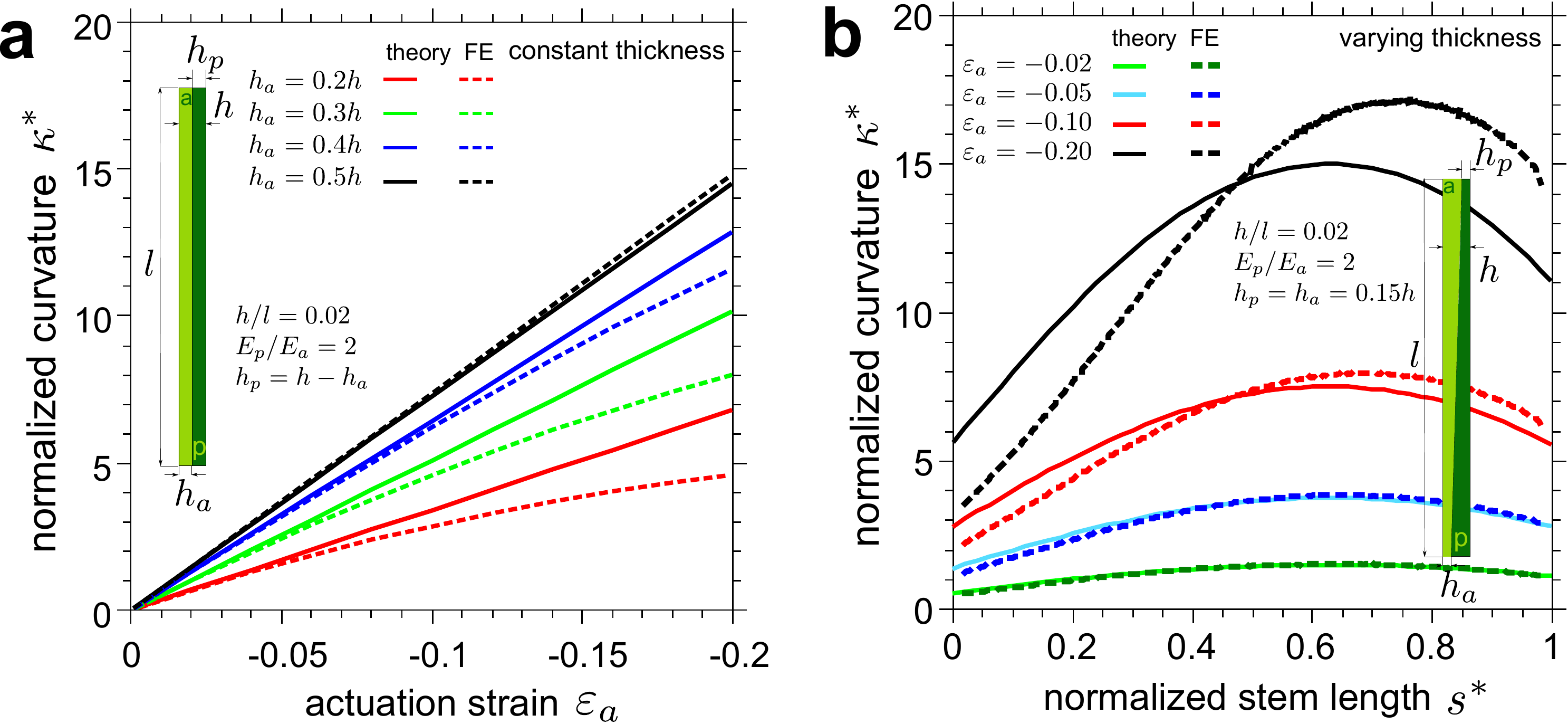}
 \caption{ Comparison between FE simulations and theoretical Timoshenko bimetallic model for normalized curvature of bilayer stems. (a) constant thickness and (b) variable thickness.}
\label{Fig_S3}
\end{figure}

In Fig.~\ref{Fig_S3}, the FE predictions of the normalized curvature of bilayer models for constant and varying thickness ratios are compared to those obtained by the Timoshenko bimetallic theory~\cite{Timoshenko25} at different actuation strains. 
For both cases, at small actuation strains, FE results are in very good agreement with theory; however, as the actuation strain increases the FE results deviates from the Timoshenko bi-metallic model which is derived based on small deformation assumption.
In FE simulations, we have taken into account geometric nonlinearities which allows nonlinear analysis of stems under large deformation, as observed in this work.
Therefore, while Timoshenko model is still a fairly good model, it is not accurate for large actuation strains. 
According to Eq.~(2) in the article, the predicted curvature of Timoshenko bi-metal model scales linearly with actuation strain. 
In contrast, FE results suggest that curvature does not magnify linearly with actuation strain and the location of maximum curvature shifts as the actuation strain increases. 
To summarize, to model the large deformation induced by stem curling upon hydration, we perform multiple simulations that account for geometric non-linearities. 
The bimetallic Timoshenko model was introduced as a limiting case, which - exact and sufficient for small deformation - still provides quite reasonable predictions for large strains.

\end{document}